\documentclass[journal,twocolumn]{IEEEtran}
\usepackage{epsfig,makeidx,color}
\usepackage{amsmath,amssymb,bbm}
\usepackage{cite,graphicx}
\usepackage{enumerate}

\def\uE{{\mathbb E}}

\DeclareMathOperator*{\argmin}{\arg\!\min}

\newtheorem{mylemma}{\bf Lemma} 
\newtheorem{corollary}{\bf Corollary} 

\def\deft{ \buildrel \triangle \over = }

\def\be{ \begin{equation} }
\def\ee{ \end{equation} }
\def\bea{ \begin{eqnarray} }
\def\eea{ \end{eqnarray} }

\def\bx{{\bf x}}

\def\bs{{\bf s}}
\def\ba{{\bf a}}

\def\bn{{\bf n}}

\def\bI{{\bf I}}

\def\bR{{\bf R}}

\def\b0{{\bf 0}}

\def\cA{{\cal A}}
\def\cC{{\cal C}}

\def\cR{{\cal R}}
\def\cI{{\cal I}}
\def\cN{{\cal N}}

\def\cV{{\cal V}}

\ifCLASSOPTIONonecolumn
  \newcommand{\figwidth}{0.50\columnwidth}
\else
  \newcommand{\figwidth}{0.90\columnwidth}
\fi

\begin{document}

\title{Re-Transmission Diversity Multiple Access
based on SIC and HARQ-IR}

\author{Jinho Choi 
\thanks{The author is with
School of Electrical Engineering and Computer Science,
Gwangju Institute of Science and Technology (GIST),
Gwangju, 61005, Korea
(\emph{Email: jchoi0114@gist.ac.kr}).
This work was supported by the GIST Research Institute
(GRI) in 2016.}}

\maketitle
\begin{abstract}
We consider multiple access with re-transmission diversity (RTxD) 
based on a hybrid automatic request (HARQ)
protocol with incremental redundancy (IR) in this paper.
In order to mitigate multiple access interference (MAI),
we employ successive interference cancellation (SIC)
at a receiver and derive conditions that all the signals from 
active users can be successfully decoded within a  
certain number of re-transmissions of IR blocks.
We consider two special cases, and based on them
we derive two multichannel random access schemes
that have sub-channels in the power and rate domains.
Through the analysis and simulations, we can show that the proposed
RTxD random access
schemes can outperform other existing similar RTxD random access schemes.
\end{abstract}

\begin{IEEEkeywords}
random access; re-transmission diversity; 
HARQ; successive interference cancellation
\end{IEEEkeywords}

\ifCLASSOPTIONonecolumn
\baselineskip 26pt
\fi

\section{Introduction}


There have been various coordinated multiple access schemes
to share a common radio resource block 
(in the frequency or time domain)
in wireless communications,
especially for uplink transmissions \cite{TseBook05}.
As opposed to coordinated multiple access,
it is often desirable to consider uncoordinated multiple access
to accommodate a number of users of low activity
with limited radio resource
for packet communications using random access schemes
\cite{BertsekasBook}. For example, slotted ALOHA \cite{Abramson77} 
and carrier sense multiple access (CSMA)
\cite{Kleinrock75} are widely studied for random access.
While random backoff plays a key role in mitigating multiple
access interference (MAI), which results in packet collision,
for most random access schemes (such as slotted ALOHA and CSMA), 
multiuser detection and other signal processing
approaches can be used to mitigate MAI when
packets from multiple users are collided
in random access.
Similar to IDMA, network-assisted diversity multiple access
(NDMA) \cite{Tsatsanis00}
employs multiuser detection with re-transmissions 
for diversity combining. 
In particular, if there are $M$ collided packets at a time,
users re-transmit the collided packets $M$ times in total
so that a receiver is able to separate them.

While re-transmission diversity\footnote{RTxD is used as the
abbreviation of re-transmission
diversity to differentiate it from repetition time diversity (RTD).}
(RTxD) 
is exploited to separate
collided signals in NDMA, it has also been employed 
in link-layer protocols for single-user transmissions. 
For example, in hybrid automatic repeat request (HARQ) protocols,
RTxD together with channel coding
plays a crucial role in providing reliable transmissions over fading channels
\cite{LinBook, WickerBook}.
HARQ can also be applied to random access \cite{Caire01},
where RTxD with incremental redundancy (IR) blocks
allows to achieve the channel capacity. It is shown in \cite{Shen09}
that the rate optimization can improve
the performance of ARQ protocols in terms of outage probability
and throughput. 
In \cite{Wu10}, 
a performance analysis of HARQ protocols is carried out under block-fading,
where closed-form expressions are derived for the throughput
and outage probability.  In \cite{Larsson14}, 
closed-form expressions are obtained for the throughputs
of various HARQ protocols including those in \cite{Shen09}.
HARQ is applied to a multiuser system with multiple
orthogonal channels in \cite{Makki14},
where resource allocation is considered with HARQ to improve
the network outage and fairness.

The notion of RTxD is also used to improve 
the probability of transmission success 
in slotted ALOHA \cite{Choudhury83}.
In conjunction with RTxD, 
signal processing approaches at a receiver such as
successive interference 
cancellation (SIC) can be used 
to suppress the MAI from collided packets \cite{Casini07, Liva11}.

In this paper, we consider multiple access
with both HARQ-IR and SIC
that can exploit RTxD and mitigate MAI, respectively.
The resulting RTxD multiple access scheme
is similar to that in \cite{Caire01} except for SIC.
It can also be seen as a generalization of that in \cite{Casini07} 
with re-transmissions of IR blocks
instead of identical coded blocks
for better performances.
The proposed scheme 
can not only take advantage of IR to have a smaller
number of re-transmissions,
but also exploit SIC to effectively mitigate MAI.
With the proposed scheme,
we find the conditions 
to guarantee successful decoding of all active signals 
within a certain number of re-transmissions
of IR blocks from active users.
While we mainly focus on the application of
the proposed scheme to random access 
in this paper, the proposed scheme can also be employed
for coordinated multiple access where conditions 
for successful decoding of all active signals 
within a certain number of re-transmissions
can be imposed by a scheduler.
In this case, we can show that the spectral efficiency
can be higher than that of orthogonal multiple access
due to SIC.

The main contributions of the paper
are as follows: 
\emph{i)} We apply HARQ-IR to 
multiple access with SIC at a receiver 
to derive an RTxD multiple access scheme,
which outperforms other existing random access schemes,
and find conditions for successful decoding within 
a certain number of re-transmissions;
\emph{ii)} From those conditions with some special cases,
we further derive two multichannel random access
schemes that have sub-channels in the power domain 
and rate domain.

The rest of the paper is organized
as follows.
In Section~\ref{S:SM},
we present the system model and 
the proposed scheme 
for RTxD multiple access with HARQ-IR and SIC.
We analyze the 
length of frame or required number of re-transmissions
for successful decoding 
and consider two special cases of the
proposed scheme in Section~\ref{S:LF}.
With multiple power levels, we propose a multichannel random access
scheme in Section~\ref{S:PDMA}.
We also derive another multichannel random access scheme 
with multiple rates in Section~\ref{S:RDMA}.
Both random access schemes are based on two special
cases studied in Section~\ref{S:LF}.
We present simulation results in 
Section~\ref{S:Sim} and conclude the paper with
some remarks in 
Section~\ref{S:Con}.

{\it Notation}:
Matrices and vectors are denoted by upper- and lower-case
boldface letters, respectively.
$\bI$ stands for the identity matrix.
$\uE[\cdot]$
and ${\rm Var}(\cdot)$
denote the statistical expectation and variance, respectively.
$\cC \cN(\ba, \bR)$
($\cN(\ba, \bR)$)
represents the distribution of
circularly symmetric complex Gaussian (CSCG)
(resp., real-valued Gaussian)
random vectors with mean vector $\ba$ and
covariance matrix $\bR$.

\section{System Model}	\label{S:SM}

In this section, we consider a system consisting of
a base station (BS) and multiple users for uplink transmissions
by sharing a common channel.
Denote by $K$ the number of users. 
We assume that a frame consists of multiple slots.
At each frame, a subgroup of users can be active
and send signals to the BS.
Each active user employs HARQ-IR.
For RTxD, an active user keeps sending IR blocks 
in every slots until the BS sends
a positive acknowledgment (ACK). 
Note that in the conventional HARQ-IR scheme,
each user may have an orthogonal radio resource block so
that there is no interference from other users.
On the other hand,
in the proposed scheme, multiple active users share a common
radio resource block for multiple access. Thus, the received signal
during time slot $t$ is given by
\be
\bx_t = \sum_{k \in \cI} h_{k} \bs_{k,t} + \bn_t,
	\label{EQ:bx}
\ee
where $h_{k}$ and $\bs_{k,t}$ represent
the channel coefficient and IR block, respectively,
from user $k \in \cI$ over time slot
$t$ to the BS, and $\bn_t \sim \cC \cN(0, \bI)$ is the background noise
(note that the noise variance is normalized for convenience).
In addition, $\cI$ denotes the index set of active users
in the current frame.
We assume that $\cI$ is invariant within a frame
and denote by $M$ the number of active users, i.e.,
$M = |\cI|$.
For convenience, throughout the paper, we assume that 
$\bs_{k,t} \sim \cC \cN (\b0, P_k \bI)$, i.e., Gaussian
codebooks are used for IR blocks, 
where $P_k$ is the transmission power of user $k$
and each IR block, $\bs_{k,t}$, is a codeword.
In addition, for convenience, we let
$\alpha_k = |h_k|^2$ and assume that
$\{\alpha_k\}$ is known at the BS.

It is clear that conventional HARQ-IR becomes a special case with
$|\cI|= 1$. That is, the above multiple access scheme 
reduces to conventional HARQ-IR if
there is only one user at a time, say user $k$. 
In this case (i.e., conventional HARQ-IR), the minimum number of IR blocks
required for successful decoding 
under the assumption of Gaussian codebooks
can be given by
\cite{Caire01, Choi13}
\be
T_{\rm CH} = \min\left\{
T \,|\, \sum_{t=1}^T \log_2 (1+ \alpha_k P_k) \ge R_k
\right\},
	\label{EQ:TCH}
\ee
where $R_k$ is the (initial) rate of user $k$
and $T$ represents the number of re-transmissions. Note
that $T_{\rm CH}$ is a 
random variable (stopping time).
As mentioned earlier, multiple users can transmit signals
using HARQ-IR without MAI 
as in \eqref{EQ:TCH}
if they are
assigned to orthogonal radio resource blocks.

At each time slot, the BS is to decode signals from users.
As usual in HARQ protocols, 
if the BS fails to decode signals,
it sends 
a negative acknowledgment (NACK). If the BS can
decode certain users' signals,
it broadcasts ACK with the indices of those users.
The users whose signals are decoded by the BS do not
send signals in the rest of the current frame, while the other active
users keep sending IR blocks.
The decoding process at the BS 
is divided into multiple stages, and in each stage,
we assume that the BS succeeds to decode signals from 
an active user with SIC.
As a result, 
we can have
\be
\cI(n) = \cI(n-1) \setminus k(n),
\ee
where $\setminus$ represents
the set difference (i.e., $B \setminus A
= \{x \in B \,|\, x \notin A\}$),
$\cI(n)$ denotes the index set of active
users whose signals are not decoded at stage $n$, and
$k(n)$ represents the index of the user whose 
signals are successfully decoded at stage $n$.
Clearly, we have $\cI {(0)}= \cI$
and the number of stages is $M$
(in this case, at each stage, 
we assume that only one user's signal is decoded).

\section{The Length of Frame}	\label{S:LF}

In this section, we study the length of frame,
denoted by $T_{\rm F}$, or the required number of 
re-transmissions of IR blocks 
from the last active users for successful decoding 
in the proposed RTxD multiple access scheme.
Clearly, $T_{\rm F}$ is a random variable 
that depends on 
the number of active users,
their channel coefficients, and signal powers.

\subsection{The Length of Frame}

The number of IR blocks required 
for stage 1 or the first successful decoding is given by
$$
\tau {(1)} = \min_{ k \in \cI(0) } \tau_k (1),
$$
where
$$
\tau_k (1) = \min \left\{T \, |\, \sum_{t=1}^T \log_2
(1 + \gamma_k(0) ) \ge R_k  \right\}, \ k \in \cI(0).
$$
Here, $\gamma_k (0)$
is the signal-to-interference-plus-noise
ratio (SINR) before SIC or at stage 0,
which is given by 
\be
\gamma_k(0) = \frac{\alpha_k P_k}
{\sum_{m \in \cI(0) \setminus k} \alpha_m P_m + 1}, \ k \in \cI(0),
	\label{EQ:SINR0}
\ee
where 
$\sum_{m \in \cI(0) \setminus k} \alpha_m P_m$ is the 
power of MAI when the signal from user $k$ is to be decoded.
As mentioned earlier, 
since $k(1)$ represents the index of the first
user whose signals can be successfully decoded at
slot $T {(1)}$ in stage 1,
we have
$k(1) = \argmin_{k \in \cI(0)} T_k (1)$.

Denote by $\bx_t (n)$ the received signal after SIC 
at stage $n$ during slot $t$.
Let $\bx_t(0) = \bx_t$.
At stage $n$, up to time 
slot $t = \tau (n)$, after SIC, $\bx_t(n)$ can be updated
from $\bx_t (n-1)$ as follows:
\begin{align}
\bx_t (n)
& = \bx_t (n-1) - h_{k(n)} \bs_{k(n),t} \cr
& = \sum_{m \in \cI(m-1) \setminus k (n)} h_{m} \bs_{m,t} + \bn_t \cr
& = \sum_{m \in \cI(m)} h_{m} \bs_{m,t} + \bn_t,
\ t = 1, \ldots, T {(n)}.
\end{align}
Note that since the BS sends ACK to user $k(n)$,
this user does not send any 
IR blocks during the rest of the current frame. That is,
there is no signal from user $k(n)$ for
$t > T {(n)}$. Thus, we can claim that
the resulting received signal at stage $n$ 
(after SIC) is given by
\be
\bx_t {(n)}
 = \sum_{m \in \cI {(n)}} h_{m} \bs_{m,t} + \bn_t
\ee
for all $t$ within the current frame.
From this,
we can define the SINR at stage $n$ as
\be
\gamma_k {(n)} = \frac{\alpha_k P_k}{\sum_{m \in \cI {(n)} \setminus k}
\alpha_m P_m + 1}, \ k \in \cI {(n)}.
\ee
The number of re-transmissions of IR blocks required for stage $n+1$ 
is now defined as
\be
\tau {(n+1)} = \min_{k \in \cI (n) } \{\tau_k (n+1)\},
	\label{EQ:TminT}
\ee
where
\be
\tau_k (n+1) = \min\left\{T \, |\, 
\sum_{t=1}^T \log_2
(1 + \gamma_k {(n)} ) \ge R_k \right\},  \ k \in \cI {(n)}.
	\label{EQ:T_kn}
\ee
In addition, we have
$k(n+1) = \argmin_{k \in \cI(n)} \tau_k (n+1)$.

For example, assume that $M = 2$.
Then, we have
$\tau(1) = \min \{\tau_1(1), \tau_2(1)\}$,
where
\begin{align*}
\tau_1 (1) & = \min\left\{T\,|\, \sum_{t=1}^T \log_2 \left(1 + 
\frac{\alpha_1 P_1}{ \alpha_2 P_2 + 1} \right) \ge R_1 \right\} \cr
\tau_2 (1) & = \min\left\{T\,|\, \sum_{t=1}^T \log_2 \left(1 + 
\frac{\alpha_2 P_2}{ \alpha_1 P_1 + 1} \right) \ge R_2 \right\}.
\end{align*}
If $\tau_1 (1) < \tau_2(2)$, user 1's signal can be decoded first
and canceled at time $\tau (1) = \tau_1 (1)$.
In this case, at stage 1, the BS sends ACK to user 1.
After SIC, the number of IR blocks for stage 2 becomes
$$
\tau(2) 
= \min\left\{T\,|\, \sum_{t=1}^T \log_2 \left(1 + 
\alpha_2 P_2 \right) \ge R_2 \right\}.
$$
Note that if $\tau(2) \le \tau(1) = \tau_1(1)$,
the BS sends ACK to user 2 too. 
Thus, at the end of slot $\tau (1)$, the BS can send
ACK to users 1 and 2 simultaneously in this case.
Otherwise,
user 2 keeps sending IR blocks until the BS sends
ACK at the end of slot $\tau (2)$ $(> \tau_1(1))$.

From \eqref{EQ:TminT},
the length of frame can be found as
\be
T_{\rm F} = \max_n \{\tau (n) \}.
	\label{EQ:TF}
\ee
If $M = 1$ (i.e., conventional HARQ-IR),
we have 
$T_{\rm F} = \tau (0) = T_{\rm CH}$.
We can also see that if
$\tau (n+1) \le \tau (n)$, 
we have
$T_{\rm F} = \tau (0)$.

For comparison purposes, we can consider
the length of frame when SIC is not used,
which is similar to NDMA \cite{Tsatsanis00} with HARQ-IR
or the approach in \cite{Caire01} where each active user
runs his/her HARQ-IR protocol independently, but no SIC is used
at the BS.
Without SIC, the SINRs remain unchanged for all stages.
Thus, the length of frame becomes
\be
T_{\rm F, no-sic}
= \max_{k \in \cI} \left\{ \tau_k (1) \right\},
	\label{EQ:ndma}
\ee
which is the maximum of the required numbers of IR blocks
for all active users.
It is easy to see that $T_{\rm F} =
T_{\rm F, no-sic}$ if $M = 1$.
Furthermore, due to SIC, in general, we can show that
$T_{\rm F} \le T_{\rm F, no-sic}$.

Note that if identical coded blocks 
(rather than IR blocks) are repeatedly transmitted as in \cite{Casini07},
\eqref{EQ:T_kn}
is modified as
\be
\hat \tau_k (n+1) = \min \left\{T \, |\, \log_2
\left(1 + \sum_{t=1}^T 
\gamma_k(n) \right) \ge R_k  \right\}. 
	\label{EQ:hTk}
\ee
Using Jensen's inequality,
we can show that
$\hat \tau_k (n+1) \ge \tau_k (n+1)$,
which results in a longer length of frame than that in
\eqref{EQ:TF}.
We may regard the approach\footnote{Note that
in \cite{Casini07}, the length of frame 
is fixed and an active user does not
transmit packets for all slots in a frame,
which are different from the proposed scheme in this section.
Furthermore, no feedback within a frame is considered.
Despite such differences,
there are some common key features such as the use of SIC and multiple
transmissions of packets from an active user within a frame.}
in \cite{Casini07} as a special case of the proposed
scheme in this paper, since
it can be obtained by replacing IR blocks with 
identical coded blocks.

\subsection{Special Cases}	\label{SS:SC}

In this subsection, we consider two special
cases of the proposed scheme.
Based on those,
we can further derive 
two variations of the proposed RTxD multiple access 
scheme in Sections~\ref{S:PDMA} and~\ref{S:RDMA}.

\subsubsection{Equalized Overall Channel Gains}

For convenience, let
$\beta_k = \alpha_k P_k$,
which is referred to as the overall channel 
(power) gain of user $k$.
Suppose that
all the overall channel gains are equalized as
\be
\beta_k = \beta, \ k \in \cI,
\ee
i.e., the power of active user $k$ is given by
$P_k = \frac{\beta}{\alpha_k}, \ k \in \cI$.

\begin{mylemma}	\label{L:1}
With equalized overall channel gains,
the length of frame 
becomes the minimum positive integer of $T$ that satisfies
\be
R_{(n)} \le T 
\log_2 \left( 1 + \frac{\beta}{ (M-n) \beta + 1} \right),
	\label{EQ:RT}
\ee
where the $R_{(n)}$'s are the ordered rates as follows:
$$
R_{(1)} \le
\ldots \le R_{(M)}.
$$
\end{mylemma}
\begin{IEEEproof}
See Appendix~\ref{A:L1}.
\end{IEEEproof}
This result demonstrates that it is possible to 
guarantee a certain length of frame
(or access delay) by assigning rates properly with equalized 
overall channel gains as in \eqref{EQ:RT}.
We will discuss a random access scheme based
on this in Section~\ref{S:RDMA} in detail.

\subsubsection{Equal Rate}

We assume that the rates are all the same, i.e.,
$R_k = R, \ k \in \cI$.
Denote by
$\beta_{(m)}$ the $m$th largest $\beta_k$ in $\cI$, i.e.,
\be
\beta_{(1)} \ge  \ldots \ge \beta_{(M)} .
	\label{EQ:bb}
\ee

\begin{mylemma}	\label{L:ER}
With equal rate,
the length of frame becomes
the minimum positive integer of $T$ that
satisfies
\be
R \le T \log_2 \left( 1 + \frac{\beta_{(n)}}{ B_{(n)}+ 1} \right),
\ n = 1, \ldots, M,
	\label{EQ:RTB}
\ee
where
$B_{(n)}$ is the MAI at stage $n$, which is given by
\be
B_{(n)} = \sum_{m = n+1}^M  \beta_{(m)}.
	\label{EQ:Bn}
\ee
\end{mylemma}
\begin{IEEEproof}
See Appendix~\ref{A:LER}.
\end{IEEEproof}

Clearly, it is possible to 
guarantee a certain length of frame
(or access delay) by assigning powers properly
for a fixed $R$.
We will discuss a random access scheme based
on the power allocation in the next section,
i.e., Section~\ref{S:PDMA}.

We can also have the following result from Lemma~\ref{L:ER}.

\begin{corollary}	\label{C:1}
If there exists a positive integer $T$ that satisfies
\be
2^{\frac{R}{T}} - 1 \le
\zeta_{(n)}
< 2^{\frac{R}{T+1}} - 1,
\ n = 1, \ldots, M,
\ee
where
\be
\zeta_{(n)} = \frac{\beta_{(n)}}{B_{(n)} + 1},
	\label{EQ:zeta}
\ee
we have
$\tau(n) = T$ for all $n$ and $T_{\rm F} = T$.
\end{corollary}

According to Corollary~\ref{C:1}, we can see that
if the SINRs lies in the region
$[2^{\frac{R}{T}} - 1, 2^{\frac{R}{T+1}} - 1)$,
all the signals can be decoded within $T$ re-transmissions.

\section{Power-Domain Multiple Access}	\label{S:PDMA}

In this section, based on Corollary~\ref{C:1},
we derive a multichannel random access 
scheme with distributed power allocation 
which suits to random access
under the following assumption.
\begin{itemize}
\item[{\bf A1)}] Each user knows his/her channel
state information (CSI), i.e., channel power
gain, $\alpha_k$.
\end{itemize}
Note that {\bf A1)} is valid 
when the coherence time is sufficiently long
(the channel gains remain unchanged within a frame)
with time division duplexing (TDD) mode.
In this case, 
each user is able to estimate his/her channel
coefficient through downlink training (i.e., a pilot signal
transmission from the BS) prior to 
the beginning of a frame based on the channel reciprocity.

\subsection{Power Allocation for Random Access}

We define multiple power levels, denoted by $\{v_l\}$, as
\be
v_l = \Gamma (V_l + 1),
\ee
where $\Gamma$ is the target SINR and
$V_l = \sum_{m= l+1}^L v_m$.
Here, $L$ is the number of power levels,
which is a design parameter, and $V_L = 0$.
Clearly, we have $v_1 > v_2 > \ldots > v_L$ for any $\Gamma > 0$.
In addition,
after some manipulations,
we can find that
\be
v_l = \Gamma (\Gamma +1)^{L-l}.
	\label{EQ:v_l}
\ee
For convenience,
let 
$\cV = \{v_1, \ldots, v_L\}$.
Define the regions for the channel (power) gain,
$\alpha_k$, as $\{\cA_l\}$ that satisfy
$\bigcup_{l=1}^L \cA_l = [0, \infty)$
and $\cA_l \cap \cA_{l^\prime} = \emptyset, \ l \ne l^\prime$.
For convenience, $\cA_l$ is called the $l$th
channel power region.
Suppose that the 
transmission power of user $k$ can be decided as follows:
\be
P_k = P(\alpha_k) = \frac{v_l}{\alpha_k}, \ \alpha_k \in \cA_l. 
	\label{EQ:DPA}
\ee
In order to lower the transmission power by taking
advantage of known CSI,
we can assume 
$\cA_1 > \cA_2 > \ldots > \cA_L$,
where the inequality of $\cA_l > \cA_{l^\prime}$
implies 
$x > y \ \mbox{for all $x \in \cA_l$ and $y \in \cA_{l^\prime}$}$.

The power allocation 
in \eqref{EQ:DPA}
is performed at each user
without any coordination by the BS. 
With this distributed power allocation, 
a random access scheme can be considered.
In this random access, a user decides the transmission 
power as in \eqref{EQ:DPA}. 
If all active
users decide different powers, all the signals
can be decoded within a certain number of re-transmissions
as follows.

\begin{mylemma}	\label{L:3}
Suppose that the power allocation in 
\eqref{EQ:DPA} is employed for all active users and
$R_k = R$ for all $k$.
If $M \le L$ and each active user's $\alpha_k$ lies in a different $\cA_l$, 
the length of frame 
is bounded as
\be
T_{\rm F} \le \Lambda (R,\Gamma) \deft \frac{R}{\log_2 (1+ \Gamma)}.  
	\label{EQ:TL3}
\ee
\end{mylemma}
\begin{IEEEproof}
See Appendix~\ref{A:L3}.
\end{IEEEproof}

The resulting random access
is referred to as power-domain 
(or power division) multiple access (PDMA)
in this paper. PDMA can be seen as a multichannel random 
access scheme with multiple sub-channels in the power-domain.
The number of sub-channels in the power-domain is $L$.

Note that as in
\eqref{EQ:TL3}, the number of re-transmissions
is bounded by $\Lambda(R,\Gamma)$ and can be shorter than this
(when the number of active users is less than $L$) by
taking advantage of HARQ-IR (i.e., if the BS 
succeeds to decode all the signals before $\Lambda(R,\Gamma)$
re-transmissions, it sends ACK signals
to all the active users and can proceed to the next frame).

To illustrate a specific power allocation
scheme, we consider the following assumption.
\begin{itemize}
\item[{\bf A2)}] 
The channel power gains are iid and they have the 
following distribution:
\be
\alpha_{k} \sim f(x) = \frac{e^{-x/\gamma}}{\gamma} , \ x \ge 0,
\ee
where $\gamma$ is the average channel power gain,
i.e., $\uE[|h_k|^2] = \uE[\alpha_k] = \gamma$. 
The resulting channels are independent Rayleigh fading channels.
\end{itemize}

\begin{figure}[thb]
\begin{center}
\includegraphics[width=\figwidth]{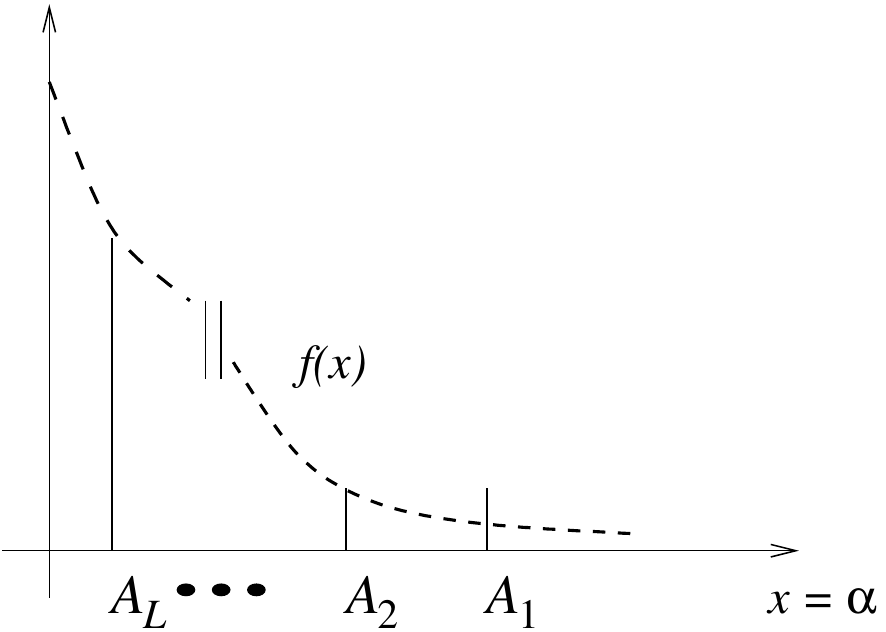} 
\end{center}
\caption{An illustration of
the division of the power domain with the $A_l$'s,
where the probability that $\alpha_k$ lies within
$[A_l, A_l-1)$ is equal for all $l$.}
        \label{Fig:div_A}
\end{figure}

Suppose that the probability that $\alpha_k$ lies in $\cA_l$
is the same for all $l$, i.e.,
\be
\Pr(\alpha_k \in \cA_l) = \frac{1}{L}.
	\label{EQ:P1L}
\ee
In this case, under {\bf A2)}, it follows
\begin{align}
\cA_l =
\left\{ 
\begin{array}{ll}
[A_1, \infty), & l = 1; \cr
[A_l, A_{l-1}), & l = 2, \ldots, L. \cr
\end{array}
\right.
\end{align}
where $A_{l-1} > A_{l}$.
If the $A_l$'s are decided 
to hold \eqref{EQ:P1L} with $A_L = 0$,
this power region assignment 
can result in a high transmission power for $\alpha_k \in \cA_L$
under {\bf A2)}.
To see  this clearly, 
we can consider the average transmission power conditioned on
$\alpha_k \in \cA_L$. From \eqref{EQ:DPA},
we have
\begin{align*}
\uE[P_k \, |\, \alpha_k \in \cA_L]
& = v_L \uE \left[\frac{1}{\alpha_k}  
\, \bigl|\, \alpha_k \in \cA_L \right] \cr
& \propto \int_0^{A_{L-1}} \frac{1}{x} e^{-x/\gamma}  dx = \infty.
\end{align*}
In order to avoid this problem, the user of the channel power
gain less than a threshold may not transmit signals.
For example, for  a given $A_L > 0$, if 
$\alpha_k < A_L$, user $k$ may not transmit signals.
In this case, under {\bf A2)}, 
the conditional pdf of $\alpha_k$ becomes
\be
f(\alpha_k \,|\, \alpha_k \ge A_L)
= \frac{C_L}{\gamma} e^{-\frac{\alpha_k}{\gamma}}, \ 
\alpha_k \ge A_L,
\ee
where $C_L = e^{\frac{A_L}{\gamma}}$.
In addition, \eqref{EQ:P1L}
is modified as
\be
\Pr(\alpha_k \in \cA_l \,|\, \alpha_k \ge A_L) = \frac{1}{L}.
	\label{EQ:PaA}
\ee
Then, after some manipulations, we can decide
the $A_l$'s as follows:
\begin{align}
A_{l-1} = A_{l} + \gamma \ln \frac{C_L L}{C_L L-e^{\frac{A_{l}}{\gamma}}}.
\end{align}
In Fig.~\ref{Fig:div_A}, we illustrate the division
of the power domain with the $A_l$'s for a given distribution
of $\alpha_k$, $f(x)$.

In PDMA, we can have the following
result for the average transmission power per user.

\begin{mylemma}	\label{L:UP}
The average transmission power per user is upper-bounded as
\begin{align}
\uE[P_k\,|\, \alpha_k \ge A_L] 
\le \frac{(\Gamma+1)^L}{L A_L}.
	\label{EQ:AUP}
\end{align}
\end{mylemma}
\begin{IEEEproof}
See Appendix~\ref{A:UP}.
\end{IEEEproof}

In \eqref{EQ:AUP}, we can see that
the average transmission power can be high
for a large $L$ as the upper-bound grows exponentially with $L$.
While it is not desirable to have a large $L$ in terms of 
the transmission power,
more signals can be successfully transmitted for a larger $L$ in PDMA.

\subsection{Outage Events}

In this subsection, we consider outage events in PDMA.
There are two different cases of outage events
that result in a longer length of frame than
$\Lambda(R,\Gamma)$ as follows:
\begin{itemize}
\item The first outage event is the event that
$M > L$. Then, there have to be 
multiple active users choosing the same
power level.
This event is referred to as overflow event.
\item The second outage event is that 
there are multiple users of the
same power levels although $M \le L$.
This event is referred to as power collision.
\end{itemize}

In order to see the performance in terms of
the probability of outage events,
we consider the following assumptions.

\begin{itemize}
\item[{\bf A3)}] A user becomes active 
independently with access probability $p_a$.
\end{itemize}

According to {\bf A3)}, 
the probability of overflow event becomes
\begin{align}
P_{{\rm out},1} 
= \Pr(M > L) 
= \sum_{m=L+1}^K \binom{K}{m} p_a^m (1-p_a)^{K-m}.
	\label{EQ:P1}
\end{align}
Thus, in order to have a low probability of overflow event,
it is desirable to have $L > K p_a$.

For the case of \eqref{EQ:PaA},
the probability of power collision events
conditioned on $M$ active users is given by
\cite{Mitz05}
\begin{align}
P_{{\rm out}, 2}  (M) 
= 1 - \prod_{i=1}^M \left(1- \frac{m}{L} \right) 
\approx \tilde P_{{\rm out}, 2} (M) 
= 1 - e^{- \frac{M^2}{2 L}},
	\label{EQ:p2a}
\end{align}
where the approximation is reasonable when $M \ll L$.

\begin{mylemma}	\label{L:P2}
Under {\bf A3)},
an upper bound on
the expectation of $\tilde P_{{\rm out}, 2} (M)$,
(i.e., the average probability of power collision) 
is given by
\begin{align}
\uE[ \tilde P_{{\rm out}, 2} (M) ]
\le 1 - e^{\uE \left[ - \frac{M^2}{2L} \right]}
= 1 - e^{- \frac{K p_a}{2L} \left( (K-1)p_a + 1\right) }.
	\label{EQ:etP}
\end{align}
\end{mylemma}
\begin{IEEEproof}
See Appendix~\ref{A:LP2}.
\end{IEEEproof}

From \eqref{EQ:etP} and \eqref{EQ:AUP},
we can observe a trade-off between the transmission power
and outage probability in terms of $L$.
That is, a larger $L$ results in a lower probability
of power collision, while a small $L$ is 
desirable to have a low transmission power.

We have a few remarks as follows.
\begin{itemize}
\item 
According to Lemma~\ref{L:3},
it is expected that all the signals from active
users can be decoded within $\Lambda(R,\Gamma)$
re-transmissions if there is no outage event.
However, there could be outage events and the BS 
may need to choose one of the following two options:
\begin{enumerate}
\item
the BS can allow the users
to transmit IR blocks until all the signals from the users
can be decoded;
\item
the BS can send the final ACK/NACK messages
after $\Lambda(R,\Gamma)$ re-transmissions and proceed to the next
frame.
\end{enumerate}
In the second option, the active users
experiencing outages become backlogged users
together with the users of low channel
gains (i.e., $\alpha_k < A_L$).
Those users (of low channel gains and
experiences outage events)
can try to transmit IR blocks in the next frame immediately.
This approach is called fast re-trial \cite{YJChoi06, Choi16},
which can only be employed in multichannel random access
schemes.
Note that fast re-trial is possible as there are 
multiple sub-channels in the power domain in PDMA.

\item While we mainly focus on random access in this section,
it is possible to exploit multiple sub-channels in the power domain
for coordinated multiple access with a scheduler to avoid
outage events.
For example, a scheduler at the BS can assign unique power levels
to up to $L$ users at the cost of signaling overhead,
which can guarantee a maximum length of frame,
$\Lambda (R,\Gamma)$.
\end{itemize}

\section{Rate-Domain Multiple Access
over Block-Fading Channels}	\label{S:RDMA}

In this section, we consider a variation
of the RTxD random access based on HARQ-IR
in Section~\ref{S:SM}
over block-fading channels that
vary from one slot to another. 
In this case, \eqref{EQ:bx}
becomes
\be
\bx_t = \sum_{k \in \cI} h_{k,t} \bs_{k,t} + \bn_t,
\ee
where $h_{k,t}$ 
is the channel coefficient from user $k$ to the BS
at time slot $t$. We 
assume that $h_{k,t}$ is independent.
Note that we do not consider Assumption {\bf A1)} in this section.
Thus, instantaneous CSI is not exploited by users.


As in \eqref{EQ:SINR0}, for fast
block-fading channels, we can have
the initial SINR at stage 0 as follows:
\be
\gamma_{k,t} (0) = \frac{\alpha_{k,t} P_k}
{\sum_{m \in \cI (0) \setminus k} \alpha_{m,k} P_m + 1},
\ee
where $\alpha_{k,t} = |h_{k,t}|^2$.
For convenience, we assume that 
$P_k$ is decided to have the same average
receive power at the BS for all $k$ as follows:
$$
\uE[\alpha_{k,t} P_k] = \gamma_k P_k 
= U, \ \mbox{for all $k$},
$$
where $\gamma_k = \uE[\alpha_{k,t}]$.
Furthermore, we consider the following assumption:
\begin{itemize}
\item[{\bf A2a)}] We assume that 
$\alpha_{k,t} P_k$ has the same distribution for all $k$,
i.e., the $\alpha_{k,t} P_k$'s are assumed to be iid.
\end{itemize}
Then, for a given $T$, the average sum rate becomes
\begin{align}
\uE \left[
\sum_{t=1}^T \log_2 \left( 1+ \gamma_{k,t} (0) \right) 
\right]
& = T \left( \psi_M (U) - \psi_{M-1} (U) \right),
\end{align}
where 
$\psi_M (x) = 
\uE\left[\log_2 \left(1+ \sum_{k=1}^M \alpha_{k,t} x \right)\right]$.
If there is an active user of a rate 
$R_{(1)} \le T \left( \psi_M (U) - \psi_{M-1} (U) \right)$,
the average number of re-transmissions for successful decoding is
equal to or less than $T$.
The signals from this user can be removed using SIC after
successful decoding.
If there is another user of a rate
$R_{(2)} \le T \left( \psi_{M-1} (U) - \psi_{M-2} (U) \right)$,
the signals from this users can be decoded and
this user'
average number of re-transmissions for successful decoding is
equal to or less than $T$. 
Based on this, we can consider a random access
scheme based on HARQ-IR in the rate domain.

Suppose that there are $L$ rates 
that are given by
\be
\eta_l = T \left( \psi_{L-l+1} (U) - \psi_{L-l} (U) 
- \delta \right)
\ l = 1, \ldots, L,
	\label{EQ:etas}
\ee
where $\delta > 0$ is a design parameter and $\psi_0 (U) = 0$.
The average sum rate (after $T$ re-transmissions) is given by
\begin{align}
\bar \eta = \frac{1}{L} \sum_{l=1}^L \eta_l 
= \frac{T}{L}(\psi_L (U) - \delta).
	\label{EQ:b_eta}
\end{align}
An active user can randomly choose one of 
the rates in $\cR = \{\eta_1, \ldots, \eta_L\}$ and transmits
IR blocks encoded at the selected initial rate.
The resulting multichannel random access scheme
is referred to as rate-domain multiple access (RDMA).

\begin{mylemma}	\label{L:R1}
In RDMA, suppose that the following two conditions hold:
\begin{align}
	\label{EQ:ieqe}
&\eta_1 < \eta_2 < \ldots < \eta_L \\
& \Pr \left( \frac{Z_l}{T} \ge \eta_l \right) > 1 - \epsilon,
	\label{EQ:1eps}
\end{align}
where
$Z_l = \log_2 \left(1+ \sum_{k=1}^{L-l+1} \alpha_{k,t} P_k \right)
- \log_2 \left(1+ \sum_{k=1}^{L-l} \alpha_{k,t} P_k \right)$.
If $M \le L$ and each active user's rate is different, 
the number of re-transmissions is equal to 
or less than $T$ with a probability higher than
or equal to $(1-\epsilon)^M$.
\end{mylemma}
\begin{IEEEproof}
See Appendix~\ref{A:R1}.
\end{IEEEproof}

For example, we may consider Rayleigh fading channels for $h_{k,t}$ as in 
{\bf A2)}. In this case,  
from \cite{Alouini99, Shin03},
we can show that
$\psi_M(x) = 
\frac{1}{\ln 2} \sum_{m=1}^M 
e^{\frac{1}{x}} E_m \left(
\frac{1}{x}
\right)$, $l = 1, \ldots, L$,
where $E_m (x) = \int_1^\infty e^{-x t} t^{-m} dt$
is the exponential integral function of order $m$.
Then, the rates in \eqref{EQ:etas}
become
\be
\eta_l = T \left(\frac{1}{\ln 2} 
e^{\frac{1}{U}} E_{L-l+1} \left(
\frac{1}{U} \right) - \delta \right).
\ee
Since
$E_m (x) > E_{m+1} (x)$,
we can claim that
$\eta_l < \eta_{l+1}$,
which implies that the condition in \eqref{EQ:ieqe} can hold
for Rayleigh fading channels.

The condition in \eqref{EQ:1eps}
is valid under Assumption {\bf A2a)}
based on the weak law of large numbers
\cite{CoverBook}. In particular, for a sufficiently large
$T$ and a small $\delta$, there might be a small $\epsilon$
satisfying \eqref{EQ:1eps}.
Consequently, for independent Rayleigh fading channels,
the signals 
from $M$ active users in 
RDMA can be decoded within $T$ re-transmissions
with a high probability if $M \le L$
and all the active users choose different initial rates.

Note that as in PDMA, there might be two outage
events. Their probabilities (and approximation) are the same as
those in \eqref{EQ:P1},
\eqref{EQ:p2a}, and \eqref{EQ:etP}.
An advantage of RDMA over PDMA is that
the transmission power may not be high 
provided that all the $\gamma_k$'s are sufficiently high. 

We can consider an asymptotic performance of RDMA
for a large $T$ to compare its achievable rate
to the achievable rate of orthogonal multiple access
with HARQ-IR as follows.

\begin{mylemma}	\label{L:R_OMA}
Suppose that $T$ is sufficiently large so that 
$\epsilon$ and $\delta$ can be sufficiently small.
Under {\bf A2a)},
the gain of RDMA over orthogonal multiple access
in terms of rate can be bounded as
\begin{align}
\frac{\bar R_{\rm rdra}}{\bar R_{\rm oma}} \le \frac{\psi_L (U)}{\psi_1 (L U)},
	\label{EQ:xoma}
\end{align}
where $\bar R_{\rm rdra}$
and $\bar R_{\rm oma}$
represent the average transmission rates of RDMA and orthogonal
multiple access (that has the number of IR blocks in \eqref{EQ:TCH}), 
respectively,
that can be achieved with the same average
transmission power.
Note that the upper-bound is achievable
if $M = L$ and there are no outage events
(i.e., if RDMA is used with a scheduler
to keep $M = L$ and to avoid choosing the same
rate by multiple users).
\end{mylemma}
\begin{IEEEproof}
See Appendix~\ref{A:R_OMA}.
\end{IEEEproof}

In \eqref{EQ:xoma}, we see that
RDMA can have a higher spectral efficiency 
than orthogonal multiple access (scheduled one without
any interference) if $\psi_L (U) \ge \psi_1 (LU)$.
The following result shows that this is true.

\begin{mylemma}	\label{L:psipsi}
Under {\bf A2a)}, we have
\be 
\psi_L (U) \ge \psi_1 (LU).
	\label{EQ:psipsi}
\ee
\end{mylemma}
\begin{IEEEproof}
See Appendix~\ref{A:psipsi}.
\end{IEEEproof}

In Fig.~\ref{Fig:xoma},
we show $\psi_L (U)$ and $\psi_1 (L U)$
for different values of $L$
when $U = 3$ dB 
for independent Rayleigh fading channels.
We can see that RDMA can provide a higher 
spectral efficiency than orthogonal multiple
access.

\begin{figure}[thb]
\begin{center}
\includegraphics[width=\figwidth]{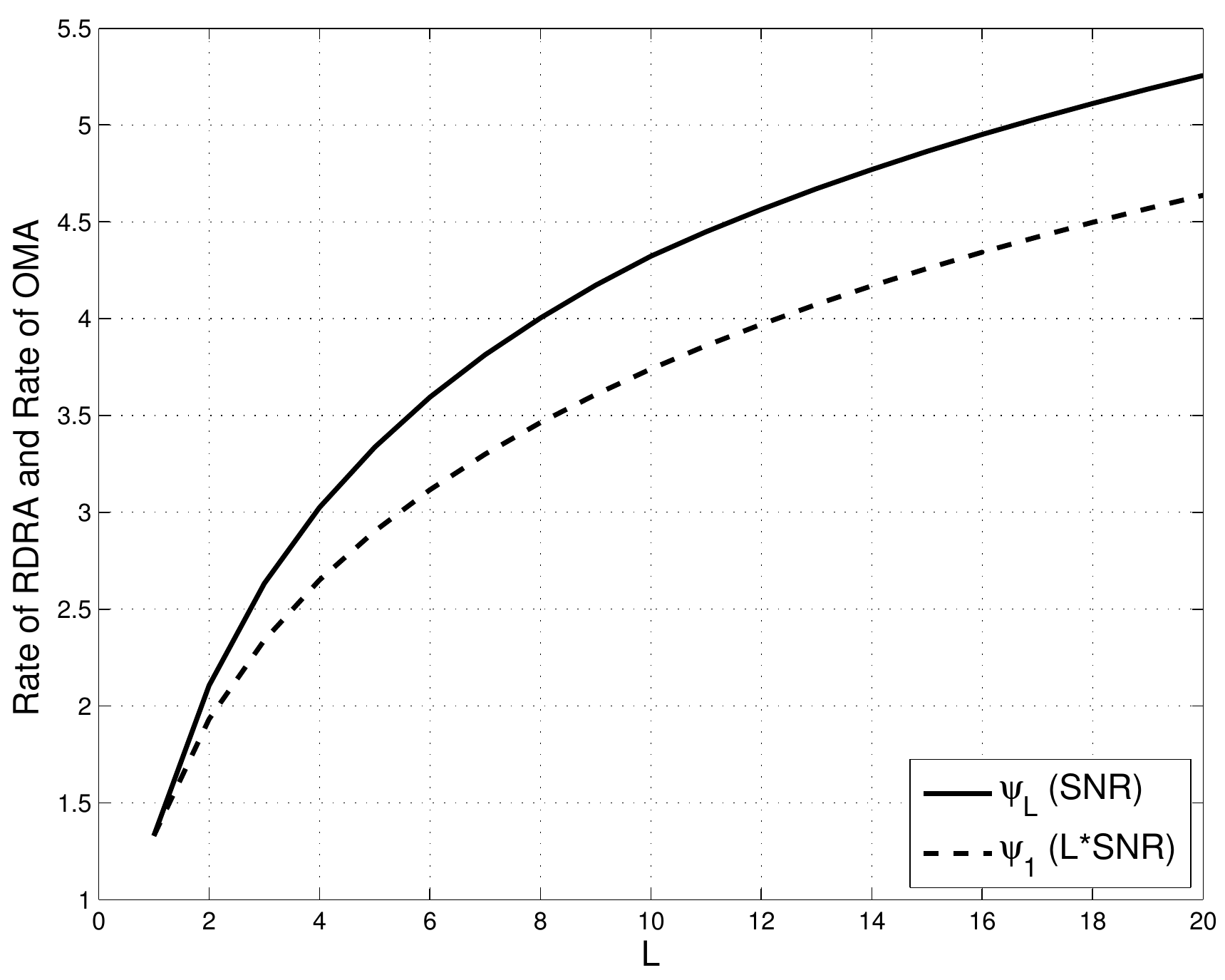} 
\end{center}
\caption{$\psi_L (U)$ and $\psi_1 (L U)$
for different values of $L$
when $U = 3$ dB for independent Rayleigh fading channels.}
        \label{Fig:xoma}
\end{figure}

We have the following remarks.
\begin{itemize}
\item If channels are invariant within a frame,
RDMA can exploit the channel reciprocity under {\bf A1)}
to take advantage of known CSI.
In this case, as in PDMA, each 
active user can choose one of the available
rates in $\cR$ according to the channel gain.
\item
PDMA can also be used for block-fading channels
where channel gains are varying for each slot.
In this case, the transmission power 
is randomly chosen from a set of pre-determined powers.
\end{itemize}

\section{Simulation Results}	\label{S:Sim}

In this section, we present simulation results
under {\bf A2)} (i.e., for independent Rayleigh fading channels).
For performance comparisons, 
we consider an RTxD random access scheme without SIC
(but HARQ-IR is employed) \cite{Caire01}, 
which has the length of frame in \eqref{EQ:ndma},
and another random access scheme 
of re-transmissions of identical coded blocks
based on \cite{Casini07},
which has the length of frame in \eqref{EQ:hTk}.
For convenience, the former scheme is referred to as
multiple access without SIC (MA without SIC)
and the latter scheme is referred to as
repetition with SIC.

Throughout this section, 
for PDMA, we assume that $A_L$ is decided 
to have $\Pr (\alpha_k < A_L) = 0.1$.
In addition, for RDMA,
$\delta$ is set to
$\delta = \frac{\eta_1}{5}$.

\begin{figure}[thb]
\begin{center}
\includegraphics[width=\figwidth]{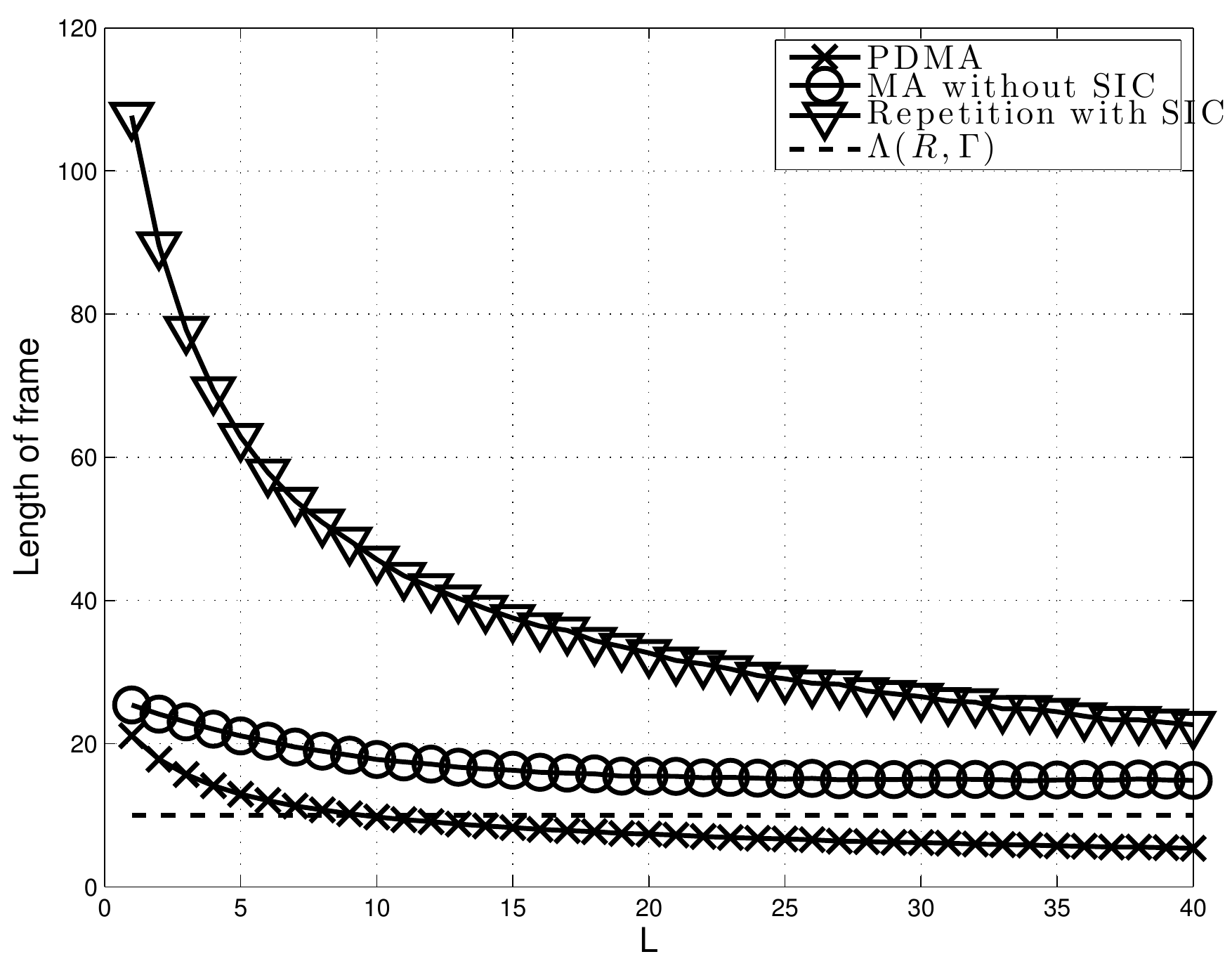} 
\end{center}
\caption{Average lengths of frame of PDMA and MA without SIC
for different values of $L$
when $\gamma = 0$ dB, $K = 50$, $p_a = 0.1$,
$R = 4$, and $\Lambda (R, \Gamma) = 10$.}
        \label{Fig:Pplt1a}
\end{figure}

\begin{figure}[thb]
\begin{center}
\includegraphics[width=\figwidth]{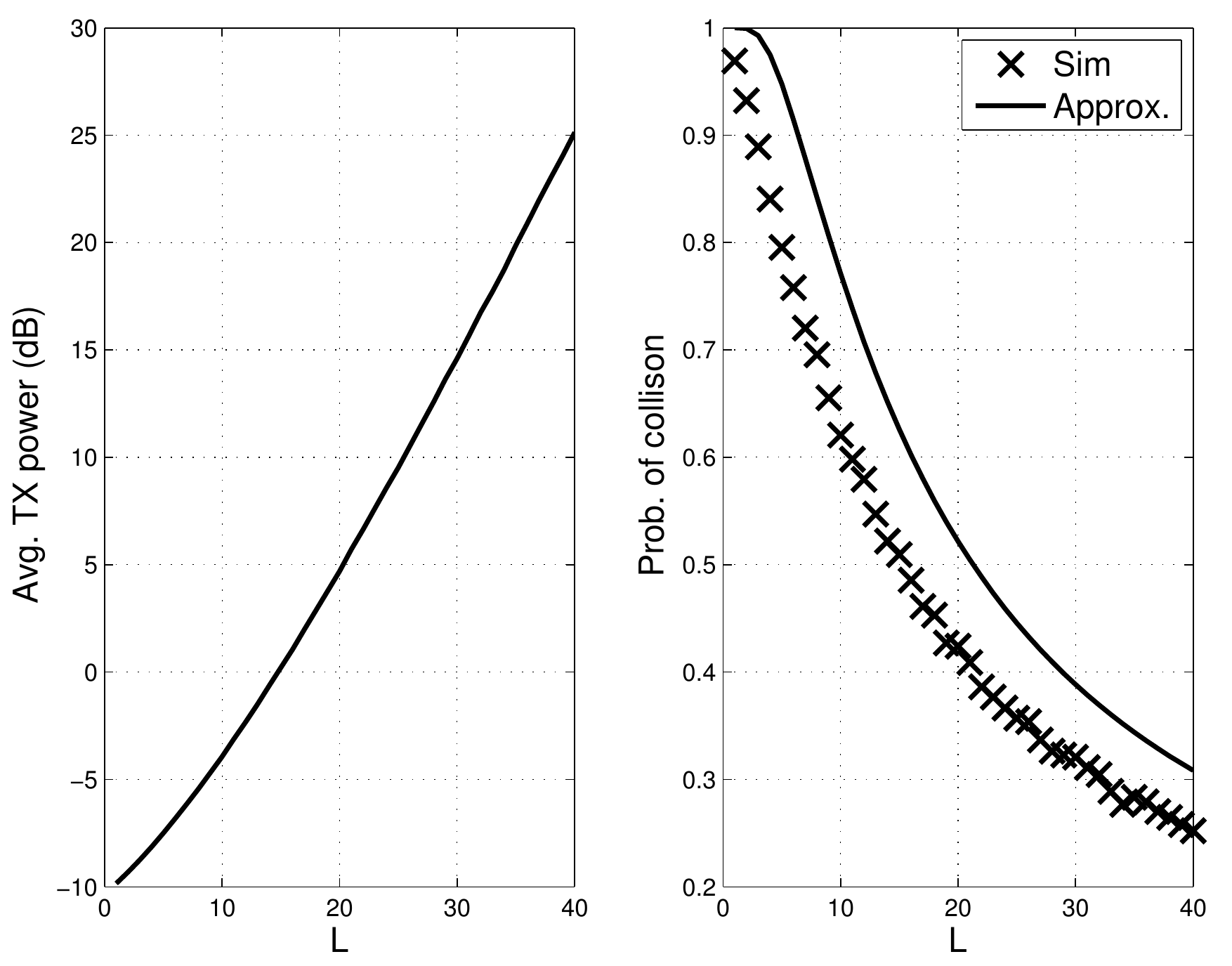}  \\
\hskip 0.5cm (a) \hskip 3.5cm (b) \\
\end{center}
\caption{Performance of PDMA for different values of $L$
when $\gamma = 0$ dB, $K = 50$, $p_a = 0.1$,
$R = 4$, and $\Lambda (R, \Gamma) = 10$:
(a) the average transmission power;
(b) the probability of power collision.}
        \label{Fig:Pplt1b}
\end{figure}

Fig.~\ref{Fig:Pplt1a}
shows the average lengths of frame of PDMA, MA without SIC,
and repetition with SIC
for different values of $L$
when $\gamma = 0$ dB, $K = 50$, $p_a = 0.1$,
$R = 4$, and $\Lambda(R, \Gamma) = 10$
(note that the value of $\Gamma$ is decided for given values of $R$
and $\Lambda(R, \Gamma)$ from \eqref{EQ:TL3}).
It is shown that PDMA can provide a smaller
length of frame than the other schemes and the lengths of frame
of all the schemes
decrease with $L$ at the cost of the transmission power
(as will be shown in Fig.~\ref{Fig:Pplt1b} (a)).
Since $\Lambda(R,\Gamma)$ is set to 10,
it is expected that the average length of frame of PDMA
is smaller than 10. If $L$ is large (i.e., $L \ge 10$),
we can see that the average length of frame of PDMA
can be smaller than 10. However, if $L$ is not large,
there might be outage events that result in a longer
length of frame than the desired one, 10.

In Fig.~\ref{Fig:Pplt1b} (a), the average
transmission power is shown for different values of $L$.
The transmission power increases with $L$,
which can be predicted by \eqref{EQ:AUP}.
As mentioned earlier, 
the lengths of frame of MA without SIC and
repetition with SIC also 
decrease with $L$ as their transmission 
powers\footnote{For simulations of 
MA without SIC, we assume 
the same transmission power as that in PDMA.}
increase with $L$.
The probability of power collision decreases with $L$
in Fig.~\ref{Fig:Pplt1b} (b)
as expected by \eqref{EQ:etP}
(we note that  \eqref{EQ:etP} is a reasonable upper-bound
on $\uE[P_{{\rm out}, 2} (M)]$).

\begin{figure}[thb]
\begin{center}
\includegraphics[width=\figwidth]{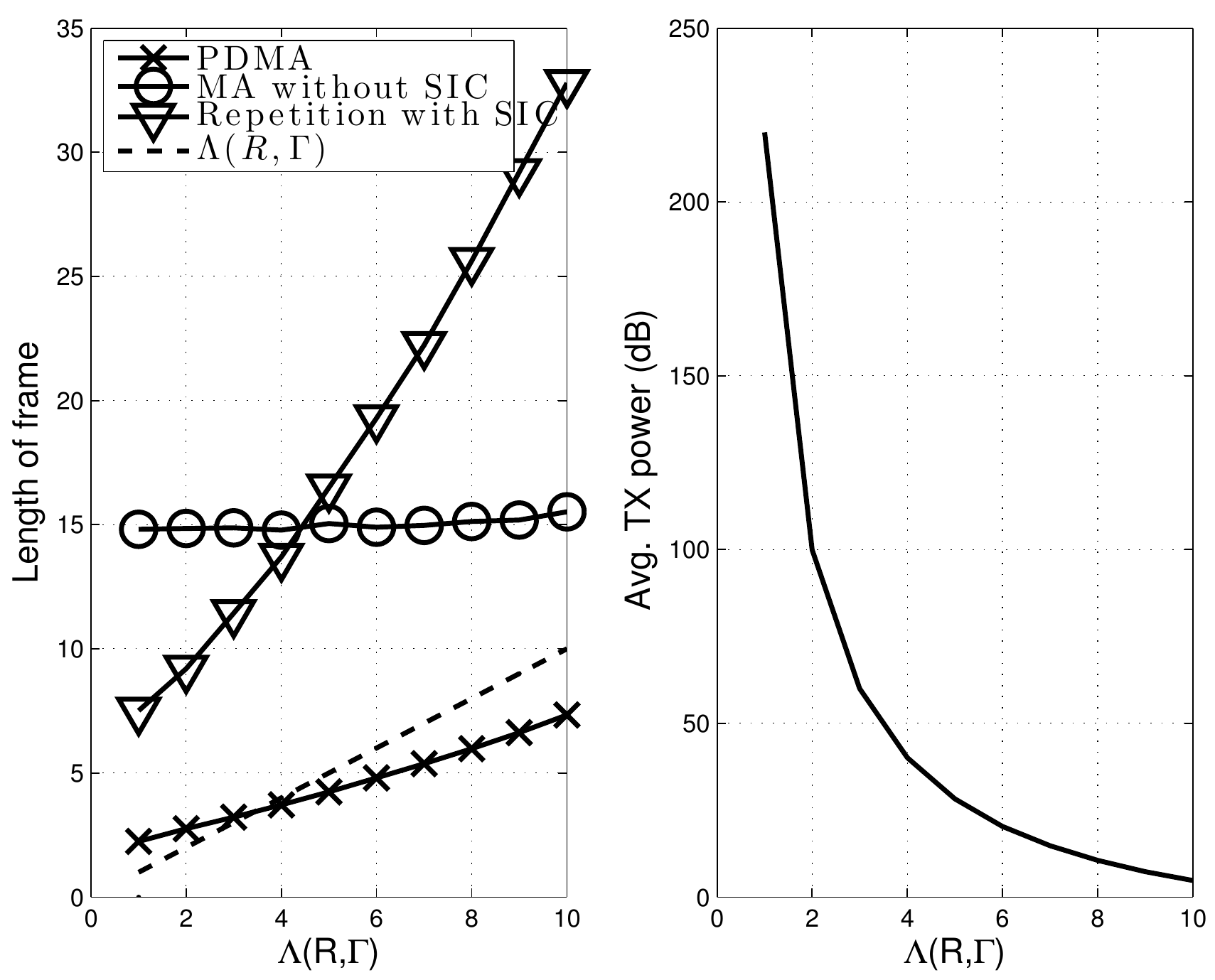}  \\
\hskip 0.5cm (a) \hskip 3.5cm (b) \\
\end{center}
\caption{Performance of PDMA for different values of $\Lambda (R,\Gamma)$
when $\gamma = 0$ dB, $K = 50$, $p_a = 0.1$,
$R = 4$, and $L = 20$:
(a) the average lengths of frame of PDMA and MA without SIC;
(b) the average transmission power.}
        \label{Fig:Pplt2}
\end{figure}

Fig.~\ref{Fig:Pplt2}
shows the performances of PDMA 
for different values of $\Lambda (R,\Gamma)$
when $\gamma = 0$ dB, $K = 50$, $p_a = 0.1$,
$R = 4$, and $L = 20$.
We can see that a large $\Lambda (R,\Gamma)$
is desirable for a smaller length of frame of PDMA. In addition,
the transmission power can be low at the cost
of a longer transmission time.

\begin{figure}[thb]
\begin{center}
\includegraphics[width=\figwidth]{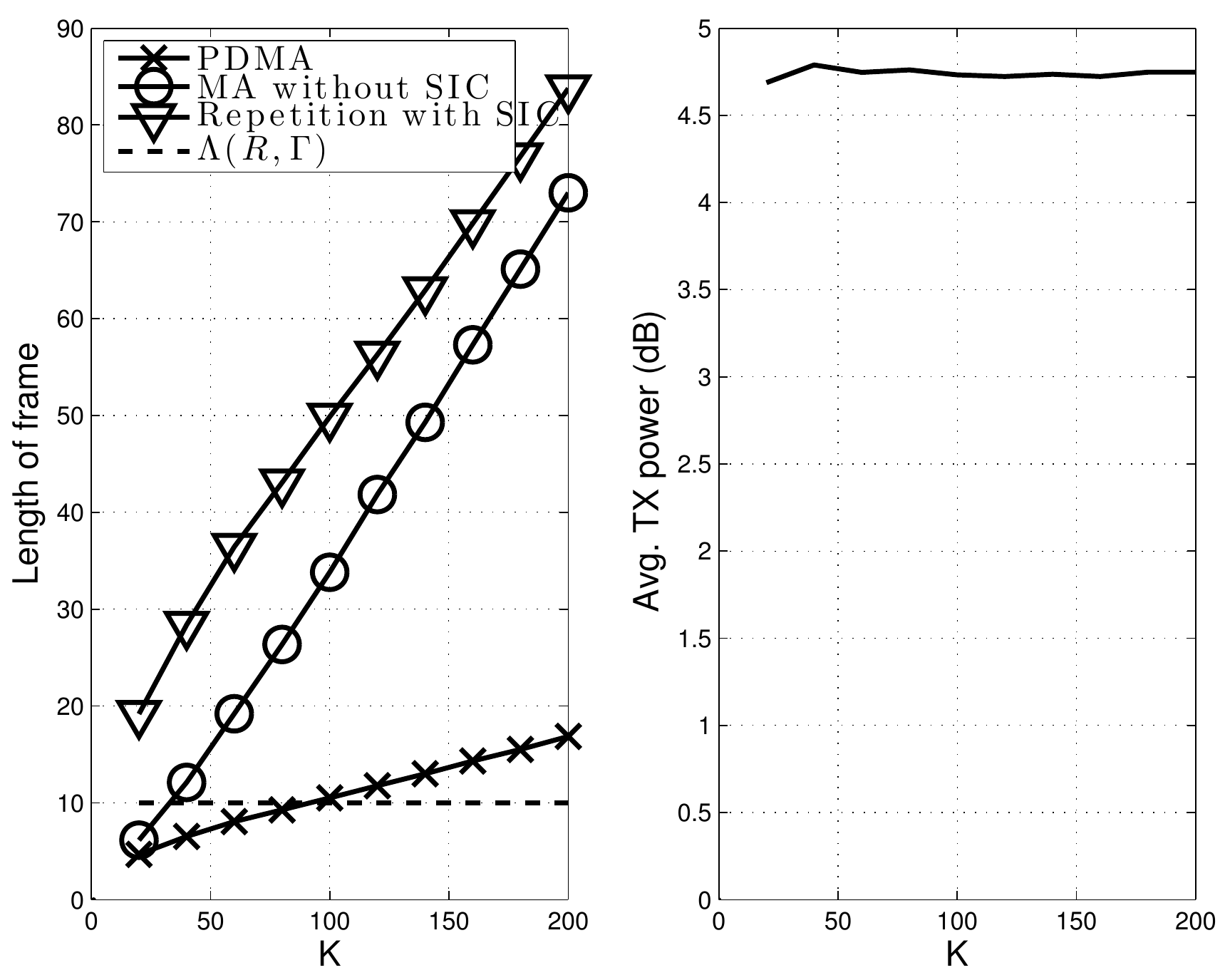}  \\
\hskip 0.5cm (a) \hskip 3.5cm (b) \\
\end{center}
\caption{Performance of PDMA for different values of $K$
when $\gamma = 0$ dB, $L = 20$, $p_a = 0.1$,
$R = 4$, and $\Lambda (R, \Gamma) = 10$:
(a) the average lengths of frame of PDMA and MA without SIC;
(b) the average transmission power.}
        \label{Fig:Pplt3}
\end{figure}

To see the impact of $K$ on the 
performances of PDMA, MA without SIC, and repetition with SIC,
different values of $K$ are considered
and simulation results
are shown in Fig.~\ref{Fig:Pplt3}
when $\gamma = 0$ dB, $L = 20$, $p_a = 0.1$,
$R = 4$, and $\Lambda (R, \Gamma) = 10$.
As $K$ increases, the average lengths of frame
of all the schemes increase as there are
more active users, while PDMA
could provide a lower length
of frame than $\Lambda (R,\Gamma)$ up to $K = 100$ users. 
Note that since $\Lambda (R,\Gamma)$ is fixed, as shown in 
Fig.~\ref{Fig:Pplt3} (b), the average transmission power
is invariant with respect to $K$.

\begin{figure}[thb]
\begin{center}
\includegraphics[width=\figwidth]{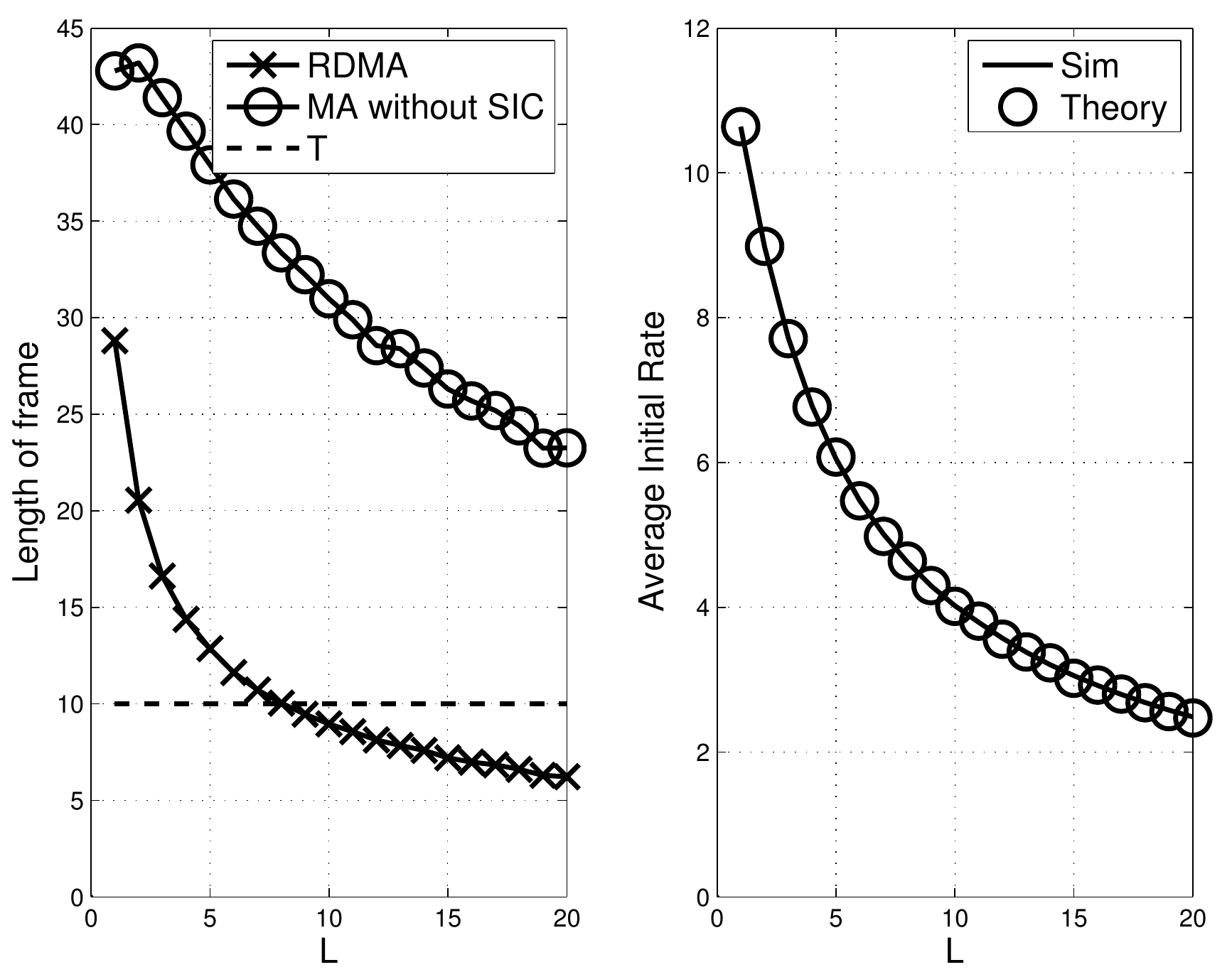}  \\
\hskip 0.5cm (a) \hskip 3.5cm (b) \\
\end{center}
\caption{Performance of RDMA for different values of $L$
when $U = 3$ dB, $K = 50$, $p_a = 0.1$, and $T = 10$:
(a) the average lengths of frame of PDMA and MA without SIC;
(b) the average transmission power.}
        \label{Fig:Rplt1}
\end{figure}

We now consider the performance of RDMA.
Fig.~\ref{Fig:Rplt1} shows 
the performances of RDMA 
and MA without SIC for different values of $L$
when $U = 3$ dB, $K = 50$, $p_a = 0.1$, and $T = 10$.
RDMA performs better than MA without SIC
in terms of the average length of 
frame as shown 
in Fig.~\ref{Fig:Rplt1} (a).
In Fig.~\ref{Fig:Rplt1} (b), it is shown that
the average initial rate decreases with $L$,
where the theoretical one is obtained from 
\eqref{EQ:b_eta}.

\begin{figure}[thb]
\begin{center}
\includegraphics[width=\figwidth]{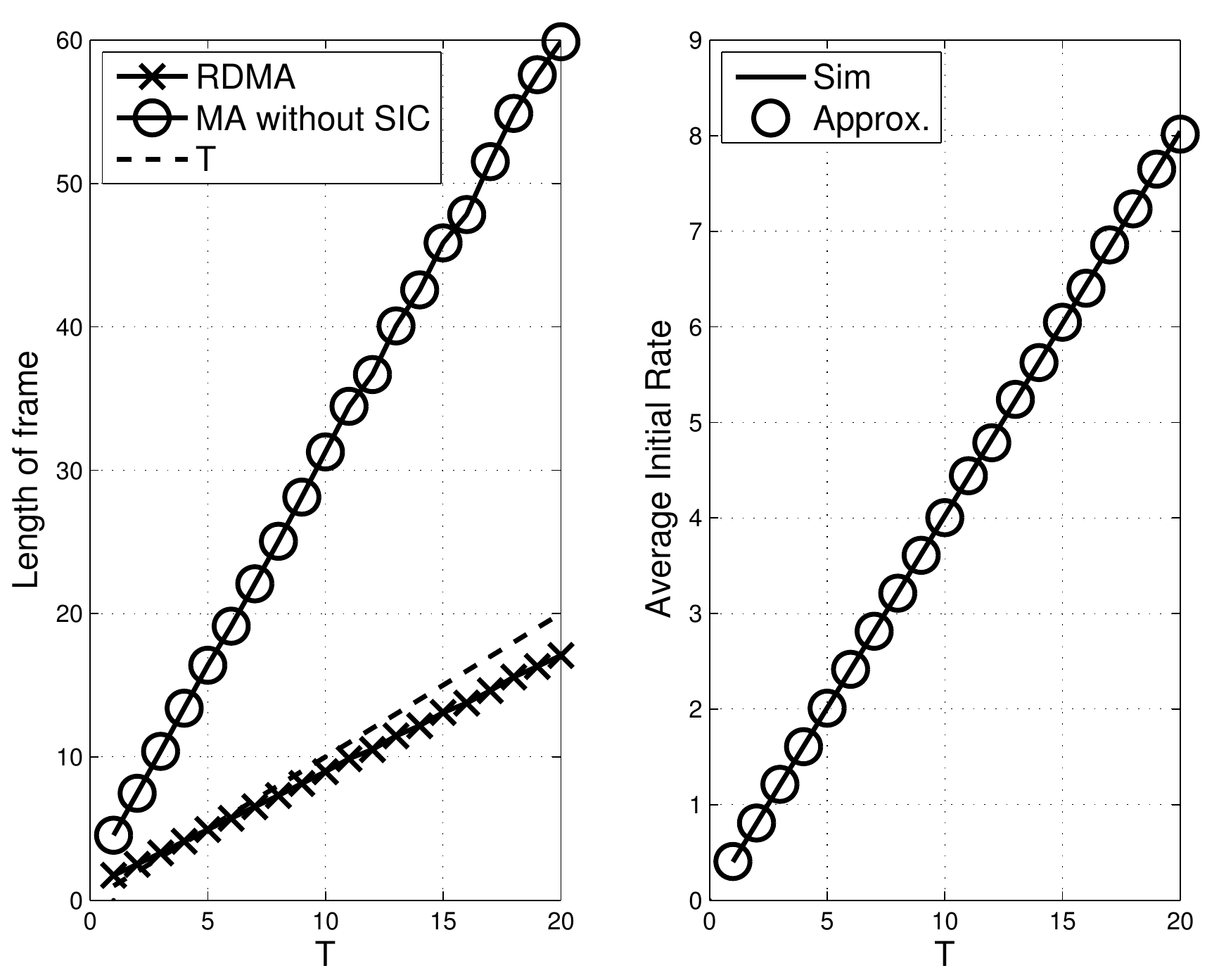}  \\
\hskip 0.5cm (a) \hskip 3.5cm (b) \\
\end{center}
\caption{Performance of RDMA for different values of $T$
when $U = 3$ dB, $K = 50$, $p_a = 0.1$, and $L = 10$:
(a) the average lengths of frame of PDMA and MA with SIC;
(b) the average transmission power.}
        \label{Fig:Rplt2}
\end{figure}

To see the impact of $T$ on the 
performances of RDMA and MA without SIC,
different values of $T$ are considered
and simulation results
are shown in Fig.~\ref{Fig:Rplt2}
when $U = 3$ dB, $K = 50$, $p_a = 0.1$, and $L = 10$.
The gap between the average lengths of frame
of RDMA and MA without SIC
increases with $T$ as shown in Fig.~\ref{Fig:Rplt2} (a), 
which shows that RDMA performs much better than MA without SIC,
especially for a large $T$.
We also see that the average length of frame of RDMA
can be smaller than $T$ for large $T$'s, which demonstrates that
RDMA can more likely guarantee that the
required number of re-transmissions 
for successful decoding
is less than or equal to $T$
when $T$ is sufficiently large.

\begin{figure}[thb]
\begin{center}
\includegraphics[width=\figwidth]{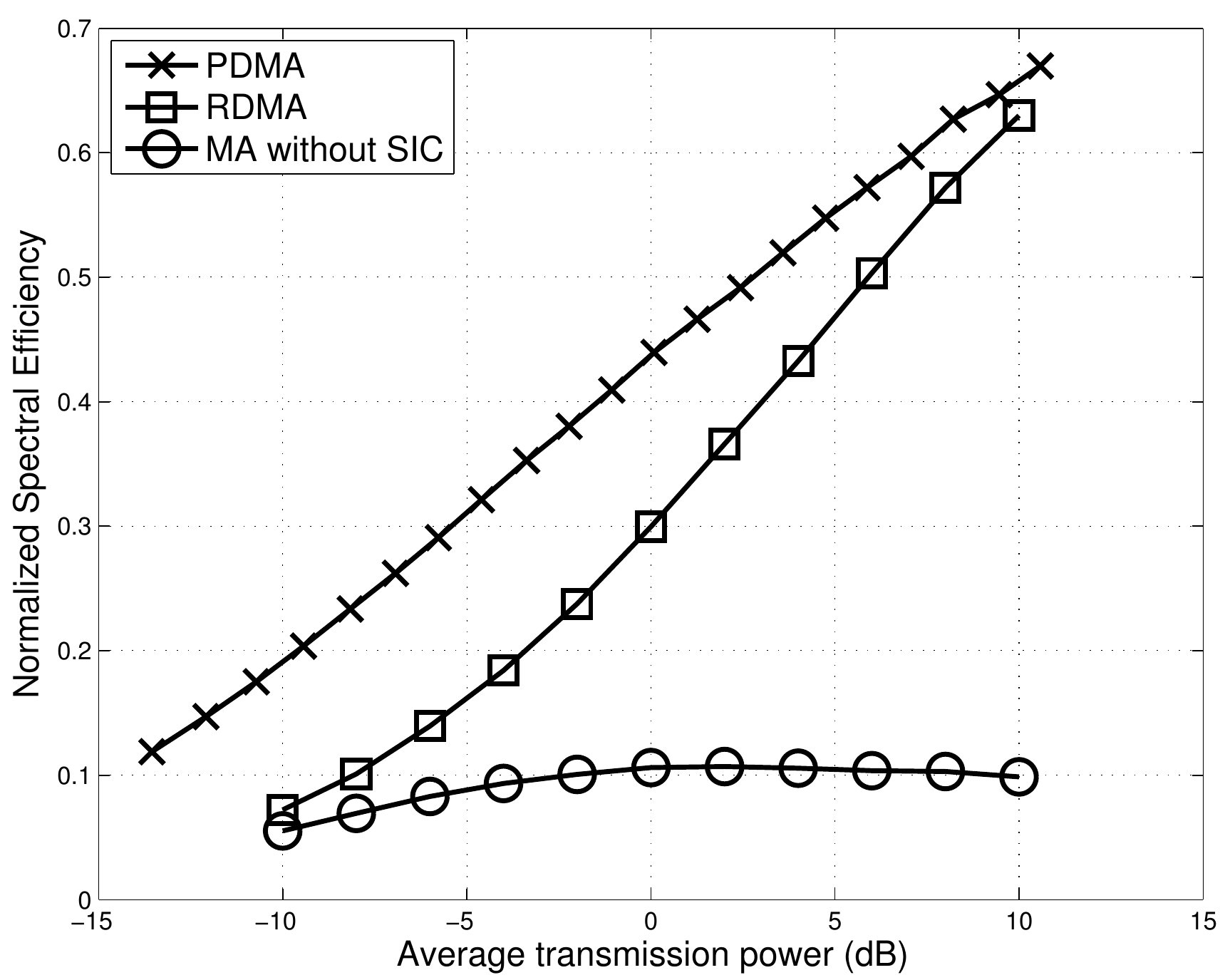} 
\end{center}
\caption{Normalized spectral efficiency versus (average) transmission
power
(with $L = 20$, $\gamma = 0$ dB, $K = 50$, $p_a = 0.1$,
and $\Lambda (R, \Gamma) = T = 10$).}
        \label{Fig:Aplt1}
\end{figure}

Finally, 
in Fig.	\ref{Fig:Aplt1},
we show the normalized spectral efficiencies
(i.e., the (average) initial rate per user over average length of frame)
of PDMA, RDMA, and MA without SIC
for different transmission powers
with $L = 20$, $\gamma = 0$ dB, $K = 50$, $p_a = 0.1$,
and $\Lambda (R, \Gamma) = T = 10$.
We can see that PDMA performs better than RDMA.
This performance gain results from the power allocation with known CSI
(in \eqref{EQ:DPA}) in PDMA.
It is interesting to note that this 
gain vanishes when the transmission power is high.
In both PDMA and RDMA, the spectral efficiency
increases with the average transmission power. This implies
that SIC can effectively mitigate interference from 
other active users.
On the other hand, in MA without SIC,
the increase of spectral efficiency 
becomes saturated when 
the transmission power increases.
This saturation indicates that the spectral efficiency
mainly depends on the MAI in MA without SIC and,
as a result, the increase of
the transmission power does not help improve the spectral efficiency
unless some interference cancellation techniques (e.g., SIC)
are employed.

\section{Concluding Remarks}	\label{S:Con}

In this paper, we studied a multiple access
scheme that are based on HARQ-IR and SIC
to exploit RTxD and mitigate MAI, respectively.
We found conditions that can guarantee a certain
number of re-transmissions in the proposed multiple access
scheme. Based on the derived conditions and two
special cases, we considered
two multichannel random access 
schemes, PDMA and RDMA, that have sub-channels in the power and rate
domains, respectively. 
The analysis results showed
that the proposed multiple access scheme can have
a shorter length of frame than other existing
similar schemes, which implies that the proposed
multiple access scheme outperforms other existing ones
in terms of the spectral efficiency.
We also demonstrated that PDMA and RDMA 
can be used for coordinated multiple access. 
In this case, RDMA can have a higher spectral efficiency
than orthogonal multiple access.

While we focused on some fundamental properties of the proposed 
multiple access schemes, 
we did not consider some possible extensions in this paper.
For example, we may consider the case that the BS is equipped with
multiple antennas (which may increase the spectral efficiency)
and a hybrid approach
of PDMA and RDMA in the future as further research topics.

\appendices

\section{Proof of Lemma~\ref{L:1}}	\label{A:L1}

The initial SINR is $\frac{\beta}{(M-1) \beta + 1}$.
The number of re-transmissions for 
stage 1 is
\begin{align*}
\tau (1) 
& = \min_k \min_T 
\left\{T \,\bigl|\, T 
\log_2 \left(1 + \frac{\beta}{(M-1) \beta + 1} \right) \le R_k\right\} \cr
& = \min_T 
\left\{T \,\bigl|\, T 
\log_2 \left(1 + \frac{\beta}{(M-1) \beta + 1} \right) \le R_{(1)}\right\} .
\end{align*}
Thus, in stage 1, the signals of rate $R_{(1)}$ 
can be removed. Subsequently, we can show that
$$
\tau (n) 
= \min_T 
\left\{T \,\bigl|\, T 
\log_2 \left(1 + \frac{\beta}{(M-n) \beta + 1} \right) \le R_{(n)}\right\} .
$$
From this, we can see that the length of frame
is the minimum positive integer of $T$ that satisfies \eqref{EQ:RT}.

\section{Proof of Lemma~\ref{L:ER}}	\label{A:LER}

From \eqref{EQ:bb}, it can be shown that
the SINRs at stage 0 are given by
\begin{align*}
\frac{\beta_{(1)}}{B- \beta_{(1)} + 1} \ge
\frac{\beta_{(2)}}{B- \beta_{(2)} + 1} 
\ge \ldots \ge \frac{\beta_{(M)}}{B-
\beta_{(M)} + 1},
\end{align*}
where $B = \sum_{m}\beta_{(m)}$,
which demonstrates that the active user who will be 
decoded first is the user that has the highest channel gain,
$\beta_{(1)}$, since the rates are the same. 
Removing the signals from this user,
we can also show that
$$
\frac{\beta_{(2)}}{(B - \beta_{(1)})- \beta_{{(2)} + 1}}
\ge \ldots \ge
\frac{\beta_{(M)}}{(B - \beta_{(1)})- \beta_{{(M)} + 1}}.
$$
From this, it is clear that the decoding order
is the same as that in \eqref{EQ:bb} (i.e., the order
of the overall channel gains).
Thus, at stage $n$, from \eqref{EQ:Bn},
the SINR of the signal to be decoded is given in \eqref{EQ:zeta}.
From this, we can see that the length of frame
is the minimum positive integer of $T$ that satisfies \eqref{EQ:RTB}.

\section{Proof of Lemma~\ref{L:3}}	\label{A:L3}

Due to the power determination in \eqref{EQ:DPA},
$\beta_k$ has to be one of $\cV$, i.e., $\beta_k \in \cV$.
In addition, since the $\alpha_k$'s
lie in different channel power regions
and $M \le L$, $\beta_k$, $k \in \cI$, should have different values.
In addition, for the ordered overall channel gains,
$\beta_{(1)} > \ \ldots \ > \beta_{(M)}$,
since $\beta_{(m)} \in \cV$, we have
$\frac{\beta_{(n)}}{B_{(n)} +1 }
\ge \Gamma$.
Note that the equality holds if
$\beta_{(1)} = v_1, \ldots, \beta_{(n)} = v_n$.
Thus, according
to Lemma~\ref{L:ER}, the length of frame, $T_{\rm F}$, is 
less than or equal to $\Lambda (R, \Gamma)$,
which completes the proof.

\section{Proof of Lemma~\ref{L:UP}}	\label{A:UP}

Since if $\alpha_k \in \cA_l$, from \eqref{EQ:DPA}, we have
the following bounds:
$\frac{v_l}{A_{l-1}} <
P_k \le \frac{v_l}{A_{l}}$.
Using the upper-bound on $P_k$ and from \eqref{EQ:PaA},
we can obtain 
the following upper-bound on
the average transmission power:
\begin{align}
\uE[P_k\,|\, \alpha_k \ge A_L] 
\le \frac{1}{L} \sum_{l=1}^L \frac{v_l}{A_{l}}.
	\label{EQ:EPU}
\end{align}
Noting that $A_L < A_{L-1} < \ldots < A_1$
and using \eqref{EQ:v_l}, it follows
$\sum_{l=1}^L \frac{v_l}{A_{l}}
\le \frac{1}{A_L} \sum_{l=1}^L v_l
=\frac{1}{A_L} \sum_{l=1}^L \Gamma (1+\Gamma)^{L-l}$.
Since
\begin{align*}
\sum_{l=1}^L (1+\Gamma)^{L-l} 
& = (1+\Gamma)^{L-1}\sum_{l=0}^{L-1} (1+\Gamma)^{-l} \cr
& \le (1+\Gamma)^{L-1} \frac{1}{1 - \frac{1}{1+\Gamma}} 
= \frac{(\Gamma+1)^L}{\Gamma},
\end{align*}
we have
$\sum_{l=1}^L \frac{v_l}{A_{l}} \le \frac{(\Gamma+1)^L}{A_L}$.
Substituting this into \eqref{EQ:EPU}, we can have \eqref{EQ:AUP}.

\section{Proof of Lemma~\ref{L:P2}}	\label{A:LP2}

Using Jensen's inequality, we can show that
\begin{align*}
\uE[ e^{- \frac{M^2}{2 L}}] 
= e^{ \ln \uE[ e^{- \frac{M^2}{2 L}}] } 
\ge e^{\uE\left[ - \frac{M^2}{2 L} \right] }
= e^{-\frac{1}{2L}\uE\left[ M^2 \right]}.
\end{align*}
Under {\bf A3)}, since $M$ is a binomial random variable,
we have $\uE[M^2] = K p_a (1-p_a) + (K p_a)^2$.
From this, we can have \eqref{EQ:etP}.

\section{Proof of Lemma~\ref{L:R1}}	\label{A:R1}
Suppose that the active user 
who has the lowest initial rate is user $k$.
It can be shown that the sum
rate of this user after $T$ re-transmissions is given by
\begin{align*}
& \sum_{t=1}^T \log_2 
\left( 1+ \left( \alpha_{k,t} + \sum_{m \in \cI \setminus k}  \alpha_{m,t}
\right) P_k \right) \cr
& -
\log_2
\left( 1+ \left( \sum_{m \in \cI \setminus k}  \alpha_{m,t}
\right) P_k \right),
\end{align*}
which has the same statistical properties
as those of $Z_{L-M+1}$ as $|\cI| = M$.
Furthermore, user $k$ has the lowest rate
among $M$ active users and all the rates are different,
the rate of user $k$ has to be
equal to or less than $\eta_{L-M+1}$,
i.e., $R_k \le \eta_{L- M+1}$.
From \eqref{EQ:1eps}, we have
\begin{align}
\Pr\left( \frac{ Z_{L-M+1}}{T} \ge R_k
\right) 
\ge
\Pr\left( \frac{ Z_{L-M+1}}{T} \ge \eta_{L-M+1}
\right)   
\ge 1 - \epsilon.
	\label{EQ:inx}
\end{align}
Thus, at time $t$ (or after $T$ re-transmissions),
the signals from user $k$ can be 
decoded 
and removed by SIC with a probability equal to or higher
than $1-\epsilon$.

Once user $k$ is removed, the user of the second lowest
initial rate,  which is less than or equal to
$\eta_{L-M+2}$, may
have $M-1$ interfering signals. The sum rate
of this user might be 
$Z_{L-M+2}$.
Thus, as in \eqref{EQ:inx}, we can show that
the signals from this user can be decoded 
and removed by SIC with a probability equal to or higher
than $1-\epsilon$.
Consequently, all the signals from $M$ active
users can be decoded within $T$ re-transmissions
with a probability equal to or higher than 
$(1-\epsilon)^M$.

\section{Proof of Lemma~\ref{L:R_OMA}}	\label{A:R_OMA}

Suppose that the number of active users in RDMA, $M$,
becomes $L$ and all users have different rates.
Then, from \eqref{EQ:etas}, under {\bf A2a)},
the average rate of RDMA per user becomes
$\bar \eta = \uE[\eta_l] 
= \frac{1}{L} \sum_{l=1}^L \eta_l 
= \frac{T}{L} (\psi_L (U) - \delta)$.
This is the maximum rate of RDMA as there 
are no outage events,
which implies that 
\be
\bar R_{\rm rdra} \le \bar  \eta.
	\label{EQ:eee}
\ee

For orthogonal multiple access,
we assume that the time slot
is divided into $L$ sub-slots and in each sub-slot,
one user can transmit signals using HARQ-IR without
any interference.
In this case, according to \cite{Caire01}, for a large $T$,
the average (achievable) rate per user 
(with $T$ re-transmissions) becomes
\begin{align}
\bar R_{\rm oma} 
= \frac{T}{L} (\uE[\log_2 (1+ \alpha_{k,t} P_k)]  -\delta) 
= \frac{T}{L} (\psi_1 (L U) - \delta)
	\label{EQ:TL1}
\end{align}
with a high probability $(\ge 1 - \epsilon)$,
where $P_k$ is decided to be $L$ times higher than that in RDMA
(i.e., $P_k = L \frac{U}{\gamma_k}$)
to have the same total power.
From \eqref{EQ:eee}
and \eqref{EQ:TL1}, 
we can have \eqref{EQ:xoma}, which completes the proof.

\section{Proof of Lemma~\ref{L:psipsi}}	\label{A:psipsi}

Consider an optimization problem as follows:
\begin{eqnarray}
& \min_{\{x_k\}} \uE[ \log_2 (1+ \sum_{k=1}^L \alpha_k x_k)] & \cr
& \mbox{subject to} \ \sum_{k=1}^L x_k = L, \ x_k \ge 0,& 
\end{eqnarray}
where the $\alpha_k$'s are iid and $\alpha_k \ge 0$ as
in {\bf A2a)}. For convenience, assume that $\uE[\alpha_k] = 1$ for
all $k$. If $x_k = U$ for all $k$, 
$\uE[ \log_2 (1+ \sum_{k=1}^L \alpha_k x_k)]$ becomes 
$\psi_L (U)$. On the other hand, if $x_1 = L$ and $x_2 = \ldots = x_L = 0$,
it becomes $\psi_1 (LU)$.

Since 
$\uE[ \log_2 (1+ \sum_{k=1}^L \alpha_k x_k)]$ 
is concave, the maximum exists.
To find the optimal solution, 
with a Lagrange multiplier, $\lambda$, we
can show that the following equality has to be 
satisfied for all $k$:
\begin{align*}
& \frac{d}{d x_k} \left(
 \uE \left[ \log_2 
\left(1+ \sum_{k=1}^L \alpha_k x_k\right) \right] 
- \lambda \sum_{k=1}^L x_k \right) \cr
& = \frac{1}{\ln 2} 
\uE \left[
\frac{\alpha_k}{
1+ \sum_{k=1}^L \alpha_k x_k} \right] - \lambda = 0.
\end{align*}
From this, since 
$\uE[ \log_2 (1+ \sum_{k=1}^L \alpha_k x_k)]$ 
is symmetric with respect to $\{x_k\}$,
we can see that
the optimal solution becomes $x_k = 1$.
Thus, $\psi_L (U)$ is the maximum, which implies \eqref{EQ:psipsi}.

\bibliographystyle{ieeetr}
\bibliography{harq}
\end{document}